\documentstyle[aps,prl,psfig,twocolumn,array]{revtex} 
\begin{document}
\draft
\title{Antiferromagnetic interlayer coupling in Fe/$c$-SiFe/Fe sandwiches and multilayers}

\author{J.M. Pruneda{$^a$}, R. Robles{$^b$}, S. Bouarab{$^c$}, J. Ferrer{$^a$} and A. Vega{$^b$}}

\address{{$^b$}Departamento de F\'{\i}sica, Facultad de Ciencias, Universidad de Oviedo,
33007 Oviedo, Spain.}

\address{{$^a$}Departamento de F\'{\i}sica Te\'orica, At\'omica, Molecular y Nuclear, Facultad 
de Ciencias.\\
Universidad de Valladolid, E-47011 Valladolid, Spain.} 

\address{{$^c$}D\'epartement de Physique,
Institut des Sciences Exactes,Universit\a'e de Tizi-Ouzou, 15000 Tizi-Ouzou, Alg\a'erie.}

\address{\begin{minipage}[t]{6.0in}
\begin{abstract}
We report the first ab-initio study of the interlayer exchange coupling in Fe/$c$-FeSi/Fe 
sandwiches and Fe/$c$-FeSi multilayers. We perform several structural studies, which 
show the stability of the CsCl arrangement seen experimentally for the spacer. We find 
antiferromagnetic coupling between the Fe slabs for spacer thicknesses smaller than about 
15 $\AA$, for both sandwiches and multilayers structures. We also study the effect of 
pinholes, interface roughness or structural misconfigurations of the spacer 
on the sign and magnitude of the 
exchange constant J. We finally show that the asymptotic behavior of J 
is determined by a flat band of the c-FeSi spacer, located at the M point in the 
Brillouin zone. 
\end{abstract} 
\end{minipage}}

\maketitle

Magnetic multilayer devices (MMD) can be roughly classified according to the conducting nature 
of the spacer material, which leads to markedly different behavior of the exchange coupling
constant between the magnetic layers, J, as a function of the thickness of the spacer, 
z \cite{Bruno}. The first class comprises those systems whose spacer is a metal. 
In this case, J(z) shows oscillatory behavior, whose period is typically of a few Angstroms; 
superimposed to these oscillations, J(z) also decays as
$1/z^2$, becoming negligible after several tens of Angstrom. The second class is composed
of those devices where the spacer is a semiconductor. Now, J(z) is frequently antiferromagnetic
(AF) and its magnitude decreases exponentially with a decay length of at most two or three
Angstrom. (Fe/$c$-FeSi) MMD stand out among all such structures as having a metallic spacer 
in {\em the brink of becoming a semiconductor}. Consequently, (Fe/$c$-FeSi) MMD might
display the characteristic features of crossover, or even critical, behavior between
two different physical regimes. Indeed, the exchange coupling constant of these devices
is always AF, but has a fairly large decay length, becoming negligible at spacer thicknesses
more proper of a metallic than of a semiconducting spacer 
\cite{Briner,Mattson,Inomata,Chaiken,Vries,Strijkers}.
The details of growth and the crystal structure of (Fe/$c$-FeSi) MMD, in addition to the unique
behavior of J(z), might also make these systems strongly attractive for manufacturers 
interested in the design of spin polarized transport devices. Indeed, the iron silicide
spacer, which possesses the CsCl structure ($c$-FeSi) \cite{Chaiken,Vries,Fullerton},
can be formed by inter-diffusion of Fe and Si slabs of appropriate thicknesses  
grown epitaxially \cite{Strijkers,Fanciulli}. LEED and AES
experiments, which have been applied successfully to study epitaxial Fe/Si/Fe (001), 
have shown that the perpendicular interlayer distance in this bcc-like structure 
remains constant at $\approx$1.43 $\AA$, very close to the values for pure bulk bcc 
Fe\cite{Strijkers}. 

Despite the large amount of experimental information, and the plausible technological
relevance of these MMD, no systematic ab-initio studies of their structural
stability, electronic structure and related magnetic properties have been performed 
up to now. Only the asymptotic region of exponentially decaying antiferromagnetic exchange 
has been qualitatively analyzed in an article by Vries {\it et al.} \cite{Vries} using both 
Bruno's ideas\cite{Bruno} and a model by Shi and coworkers \cite{Shi}. We report in this Letter
a thorough study of the structural stability, the exchange constant J, and the magnetic moment
distribution of Fe/$c-$FeSi/Fe(001) trilayer and Fe/$c-$FeSi multilayer devices
as a function of the spacer thickness z. We show how interstitial defects such as 
pinholes or interface roughness affect the sign and magnitude of J for thin spacers, by 
studying several mixed 
interfaces. We have extended Bruno's theory \cite{Bruno} to account for the effect of
different effective masses on the exchange constant. This has allowed us to make an
excellent fit to the asymptotic behavior of J, which shows that the relevant contribution to the
exchange constant comes from a band in the c-FeSi spacer located at the M point of the Brillouin
zone, and not at the X point, as speculated previously \cite{Vries,Shi}.

We have determined the spin-polarized electronic structure
using a scalar-relativistic version of the $k$-space TB-LMTO method \cite{Andersen} 
developed in the atomic spheres approximation. We have used two different versions of the 
generalized gradient approximation (GGA) for the exchange and correlation potential: 
the Perdew-Wang (PW) \cite{PW} and the Langreth-Mehl-Hu (LMH) \cite{Langreth}, and benchmark 
them against preliminary results obtained in the Local Spin Density Approximation (LSDA)
\cite{Ecoss}. We model the system according to the existing LEED structural analysis of 
MBE-grown Fe/SiFe/Fe (001) sandwiches briefly mentioned above \cite{Strijkers,Fanciulli}. 
We have performed our calculations 
using both the experimental interlayer distance ($1.43 \AA$) and the lattice parameters which 
come from a total energy minimization of Fe and $c$-FeSi bulk materials, and obtained overall 
qualitative agreement between both sets of calculations. For the sake of clarity, we discuss 
here the 
results obtained using the experimental distance, unless otherwise stated. We employ the 
super-cell technique, both for trilayer (TD) and multilayer devices. For trilayer structures, 
we repeat the sequence Fe$_7$/$c$-Fe$_{(n-1)}$Si$_n$/Fe$_7$ in the (001) direction, 
with $n$ = 1 - 6. Two successive trilayer sandwiches are then separated by enough layers 
of atomic 
empty spheres to have each individual trilayer decoupled from the rest.
We have also investigated the effect of the thickness of the Fe substrate
on the stability of the interlayer exchange coupling, and found that the
results are not altered for Fe slabs thicker than the selected case of seven layers. 
The electronic and magnetic structures are calculated using an increasing number of 
$k$-points in the irreducible Brillouin zone. Convergence is obtained for 
90 $k$-points, results for 132 $k$-points being quantitatively the same.

\begin{figure}
\psfig{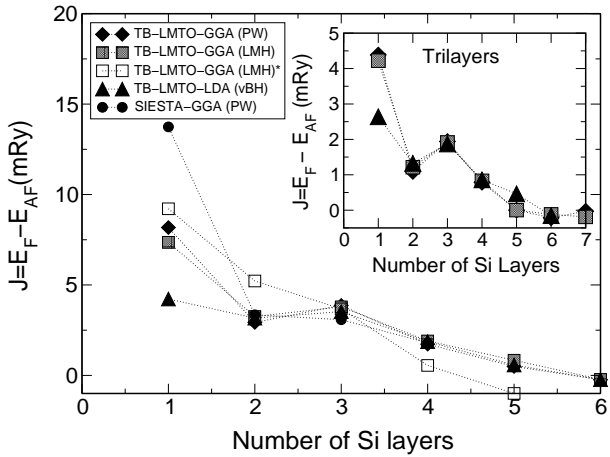}
\vspace{0.3cm}
\caption[]{Total energy difference J = E$_{F}$ - E$_{AF}$ as function 
of the number $n$ of Si atomic monolayers in the spacer for 
Fe$_7$/$c-$Fe$_{(n-1)}$Si$_{n}$/Fe$_7$(001) multilayer devices. The inset shows
J for trilayers. Open squares denote values of J obtained using the theoretical
lattice constant and the LMH functional.}
\end{figure}

We have first made structural analyses to test that the experimental result
of inter-diffusion of a thin Si film into the Fe slabs, that creates the $c$-FeSi structure, 
can be understood in terms of the energetics of the different plausible 
atomic arrangements in the spacer. We have therefore computed the total energy for  
Fe$_7$/Si$_3$Fe$_2$/Fe$_7$ TD with different layer configurations in the spacer, which 
modify the $c$-CsCl structure. We have checked, for instance, that the 
Fe$_7$/Si/Fe/Si$_2$/Fe/Fe$_7$ arrangement is 300 mRy higher in energy than the 
Fe$_7$/Si/Fe/Si/Fe/Si/Fe$_7$ configuration, and that in the former case the F 
alignment is slightly more stable than the AF one.  

We have further performed molecular dynamics simulations, relaxing the atomic positions 
in the z direction, for Fe$_9$/$c$-(Fe$_1$Si$_2$)/Fe$_9$ and Fe$_7$/$c$-(Fe$_2$Si$_3$)/Fe$_7$ 
TD using the SIESTA code. SIESTA is a Density Functional Theory code based in
pseudopotentials which allows structure 
optimizations of the whole system which are beyond the capabilities of the all-electron
TB-LMTO method, particularly when surfaces are present. Besides, it has been 
successfully applied to several magnetic structures composed of iron atoms in 
different environments\cite{gorka}. We have used 
a minimal basis set and the PW functional for exchange-correlation \cite{PW}.  
Pseudopotentials were generated with the Troullier-Martins method \cite{TM}, with 
$4s^{1}3d^{7}$ and $3s^{2}3p^{2}$ valence configurations for Fe and Si, respectively.
We have found that the forces in the theoretical bcc-like configuration just make the 
interstitial Fe atoms move slightly into the FeSi spacer, and expand a bit the Si-Si 
distances, the structure being very stable otherwise.  The obtained exchange constants J  
are also similar within 10 \% to those obtained with the TB-LMTO method in the unrelaxed 
structure, therefore lending further support to our theoretical study.   

We pass on now to discuss the behavior of J(z), defined as the difference between the 
total energies of F and AF alignments of the Fe slabs, for thin TD with perfect interfaces. 
Fig. 1 shows  the exchange constant as a function of the number n of Si atomic monolayers,
which is a measure of the thickness of the spacer.
We always find a positive J of the order of some mRy, which decreases with z in a 
non-monotonic fashion, has a bump for n = 3 and vanishes for n $\approx$ 5 - 6.
The bump might correlate with a small protuberance which is seen experimentally at 
exactly the same spacer thickness \cite{Vries}. It is not an artifact produced by the 
TB-LMTO code because we also find it when we use SIESTA. Its height depends, on the 
other hand, on the choice of the interlayer distances, becoming smaller when we use 
theoretical or relaxed values for them (see the inset in Fig. 1). The experimental peak 
seen for n=4 
seems to be somewhat sample dependent \cite{Vries_thesis}, and we find no traces of it 
in our simulations. The exchange constant becomes negligible for spacer thicknesses 
larger than five or six atomic Si layers, which 
correspond to a distance of about 15 $\AA$. Because the error bars coming from our 
calculations are of the order of the values obtained, we can only honestly 
conclude that J does not have any large oscillation for such thicknesses. Experiments 
actually show that for z larger than $\approx$13$\AA$, J(z) follows the asymptotic 
behavior of a semiconductor, with a large decay length of about 3.6$\AA$.
This overall theoretical behavior for thin trilayer devices, that was also obtained 
within the LSDA \cite{Ecoss}, is therefore in nice qualitative 
agreement with the experimental results by Vries {\it et al.}\cite{Vries}. 

We find that the exchange constant obtained for multilayers follows the same
trend as that found for trilayer structures, its magnitude being roughly a factor of two larger
(see inset of Fig 1). 
This fact can be qualitatively explained in terms of simple Heisenberg-like physics, making use
of an analogy with spin chains: the energy to flip a spin in a molecule composed of two atoms 
is half that 
required to do so in an infinite chain. We also find that for given z, J slightly oscillates 
as a function of the thickness of the iron slabs, as seen in Co/Cu MMD \cite{Bloemen,Bruno}. 

\begin{figure}
\psfig{figure=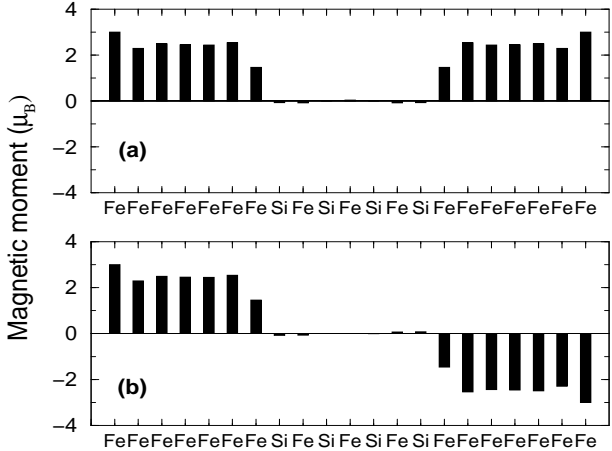,height=6cm,width=8cm,angle=-90}
\vspace{0.3cm}
\caption{Distribution of magnetic moments for 
Fe$_7$/c-Fe$_{3}$Si$_{4}$/Fe$_7$ (001) 
sandwiches in the (a) ferromagnetic and (b) antiferromagnetic configurations.
Black bars correspond to the magnetic moments of Fe whereas white 
bars correspond to Si moments.}      
\end{figure}  
Vries and coworkers observe F-type contributions superimposed on top of 
the AF-like behavior in their Kerr hysteresis loops for TD with spacer thicknesses 
smaller than about 6-7 $\AA$ \cite{Vries}. We present now results for several 
TD where we simulate a variety of atomic misconfigurations inside the spacer or at its 
interfaces with the iron slabs, because we wish to test whether they can induce 
such ferromagnetic coupling. The first one consists of the Fe$_7$/Si/Fe/Si$_2$/Fe/Fe$_7$ 
TD which was introduced a few paragraphs before. It is supposed in this case that the 
Si layers have not diffused 
completely into the Fe slabs to form the $c$-FeSi structure. We find that J is ferromagnetic, 
with J=-3.49 mRy. We present next the case of dense arrays of thin pinholes
or rough interfaces for TD with n = 1, 2, and 3 (e.g.: up to about 8 $\AA$). We simulate 
them by doubling the cross section of our unit cell, which now contains two atoms per 
atomic plane. For n=1, we exchange one atom in the Si layer by another in the last 
layer of one of the iron slabs. We create this way a periodic array of thin pinholes which 
leads to a strongly ferromagnetic J of -73 mRy (about thirty times larger in magnitude 
than the corresponding one for a pure $c$-FeSi spacer). We 
subsequently look at two different configurations for n=2. In the first one, the Fe 
and Si atoms are arranged again in such a way that there are thin bridges 
along the spacer as before. The exchange constant is then also negative, with a value 
of -33 mRy. We make the second configuration such that one of the Si layers has no iron atoms, 
which means that one of the interfaces is rough while the other one is perfect.  We find in 
this largely asymmetric case that J is equal to -0.7 mRy, which is of the same order of 
magnitude than its antiferromagnetic 
counterpart. We finally present for n = 3 the case where both interfaces are rough while
the central Si layer has no defects. We find that the exchange constant is again largely
ferromagnetic, J = -5.1 mRy. We therefore conclude that atomic misconfigurations generically 
give rise to very strong ferromagnetic couplings. A comparison of our results with those 
presented by Vries \cite{Vries} lets us infere some further conclusions about the structure
of these sandwiches, namely: (a) it is very likely that for TD with thicknesses 
smaller than about 3 $\AA$ there is a process of diffusion of iron into the Si layer 
(or vice versa), so that the Si layer is disrupted; (b) for TD with spacer 
thicknesses larger than about 4 $\AA$, diffusion
takes place but leads to the formation of a $c$-FeSi spacer with interfaces of high quality. 

We come to comment now on the profiles that we find for the magnetic moments, which we show in 
Fig. 2 for TD  with n=4. First, we have that the $c$-FeSi spacer displays tiny 
magnetic moments of order 0.05 $\mu_B$ at the interface with the iron slabs, with whom they 
couple antiferromagnetically. Second, the absolute values of these magnetic moments are almost 
identical for both alignments. Third, iron atoms in the vicinity of the interface with the spacer 
have reduced magnetic moments ($\sim 1.46 \mu_B$) due to the hybridization of their orbitals 
with those belonging to c-FeSi atoms. Surface effects, on the other 
hand, induce enhanced magnetic moments at the Fe external layers, (M$\sim 2.97 \mu_B$). 
Finally, magnetic moments at the center of the Fe slabs slightly oscillate around the bulk 
value ($\sim 2.20 \mu_B$). We find similar behavior for multilayer devices, apart from the 
obvious fact that no surface effects exist for them. 

\begin{figure}
\psfig{figure=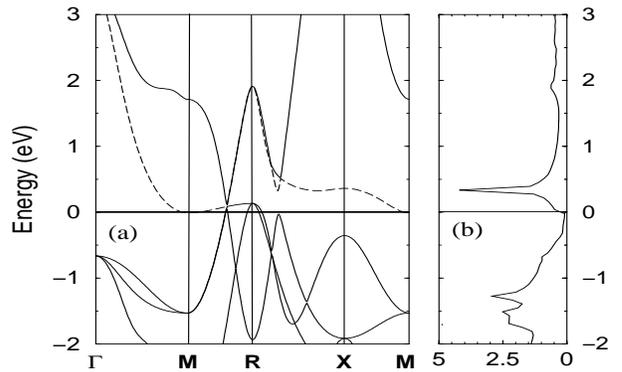,height=5cm,width=8cm,angle=-90}
\vspace{0.2cm}
\caption{(a) Details of the band structure of bulk $c$-FeSi. The band drawn with a dashed line 
has a band edge at the M point and a large flat section around the X point; (b) Total density 
of states of bulk $c$-FeSi (in states/eV), showing the band edge at the Fermi level and a sharp peak 
located about 0.3 eV above it.}      
\label{fig3}
\end{figure}  

The last part of this letter is devoted to study the asymptotic behavior of J(z). We have 
extended Bruno's theory to account for the different effective masses of electrons located 
at the relevant bands of the spacer and of the majority and minority populations of the 
iron slabs, and used it to study the 
contribution to J(z) coming from the high-symmetry points in the Brillouin zone. 
Several authors \cite{Vries,Shi} have speculated that the main contribution to J(z) should come 
from a band 
centered at the X point, which gives rise to a sharp peak in the density of states, see Fig. 
3. We have generated the band structure of bulk ferromagnetic bcc iron and c-FeSi using 
the TB-LMTO code 
and used them to fit such a band, as well as those in the iron slabs closest to the Fermi 
energy, to parabolae. This has allowed us to extract band edges $E_g$ and effective masses 
$m^*$, which we have plugged into our extension of Bruno's formula for J(z). We have found 
that the exchange constant decays in an oscillatory fashion. This behavior, which is due to 
the negative effective mass of the relevant band at the spacer, can not be 
easily reconciled with the experimental
data. We have therefore turned our attention to the contribution of the 
same band at the M point, which now gives rise to a three-dimensional band edge right at the 
Fermi energy. The naive estimate for decay length $\lambda$ would in this case be infinity. 
If such an hypothesis was confirmed, 
the exchange constant would decay as a power law, displaying therefore the typical behavior 
of critical phenomena \cite{Amit}. We nevertheless find that additional contributions to  
$\lambda$ come from the fact that the $E_g$'s of the bands at the iron slabs are not placed 
at the Fermi energy, a fact which gives rise to finite effective barriers 
(we only find power law decay when one of these edges is finely tuned to zero energy). 
We have fitted 
again the $E_g$'s and $m^*$'s of the relevant bands at the M point and introduced them into 
the formula for J(z). We have extracted this way a decay length of 3.6 $\AA$ and more 
generically an 
overall behavior in excellent agreement with the experimental results available to us. 
Fig. 4 shows how J(z) fits the exponential $e^{-z/3.6}$. The layout of this figure 
purposely mimics that of Fig. 5 in Ref. \cite{Vries} with which should be compared.
We wish to stress that all parameters entering into the expression 
for J(z) have been obtained independently from a band structure calculation, which has left 
us no room for fitting. 

\begin{figure}
\psfig{figure=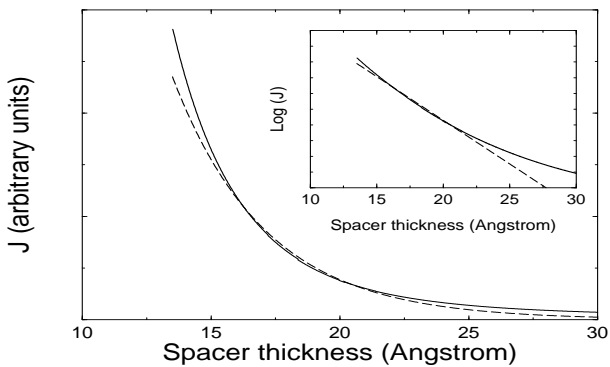,height=5cm,width=8cm,angle=-90}
\vspace{0.1cm}
\caption{
Asymptotic behavior of the exchange constant as a function of the thickness of the 
spacer. Solid line is the result from our amended version of Bruno's theory; dashed 
line is a fit to $e^{-z/3.6}$. The inset shows the logarithm of J again as a function of z. 
}      
\label{fig4}
\end{figure}

In conclusion, we have performed a thorough study of the structural stability and magnetic
properties of Fe/$c$-FeSi magnetic trilayer and multilayer devices, using a combination of density
functional methods. We have studied how dense arrays of pinholes and interface roughness, or
other structural misconfigurations affect the AFM character of the exchange constant. We have
deduced that the process of diffusion which is used to grow these MMD is very effective and
leads to interfaces of very high quality. We have made use of a slight modification of
the theory by Bruno to show how it is the M point which governs the behavior of J(z) for large
thicknesses of the c-FeSi spacer.

This work was supported by Acciones Integradas (HF1999-0041), the
spanish DGESIC (Projects PB98-0368-C02 and PB96-0080-C02) and Junta de 
Castilla y L\'eon (Grant VA 70/99). R. Robles and J.M. Pruneda acknowledge 
FPU and FPI grants from the Spanish Council of Education and Culture. We 
would like to thank Professors J. M. Alameda and C. Demangeat for fruitful 
discussions.

\end{document}